\documentclass[aps,pra,lengthcheck,preprintnumbers]{revtex4-1}
\usepackage{amsfonts}
\usepackage{amsmath}
\usepackage{amssymb}
\usepackage{dsfont}
\usepackage{graphicx}
\usepackage{color}
\usepackage[unicode]{hyperref}
\usepackage{mathrsfs}
\hypersetup{
	pdftitle={spin currents},
	pdfauthor={},
	pdfsubject={},
	colorlinks=true,
	linkcolor=blue,
	citecolor=blue,
	filecolor=black,
	urlcolor=blue
}

\begin{document}
	\title{Spin effects in the particle current of Bose-Einstein condensates\\ in synthetic gauge fields}
	\author{Hugo Bencomo Mart\' in}
	\affiliation{Departamento de F\'isica, Universidad de La Laguna, Tenerife 38200, Spain}
	\author{Antonio Muñoz Mateo}
	\affiliation{Departamento de F\'isica, Universidad de La Laguna, Tenerife 38200, Spain}
	\affiliation{Instituto Universitario de Estudios Avanzados en Física Atómica, Molecular y Fotónica (IUDEA), Universidad de La Laguna, Tenerife 38203, Spain		}
	\date{\today}
	\begin{abstract}
		Spin reveals the rotational features of a system independently of its state of motion. Hence, it is not a relativistic quantity, despite giving rise to effects that emerge naturally from relativistic equations only. Within a non-relativistic framework these effects are not inherently captured, and have to be added by resorting to symmetry principles or to the asymptotic limit of the corresponding relativistic framework. We report on one of these effects: the divergenceless spin term in the particle current density of spinor Bose-Einstein condensates. It could be observed in the setup of a synthetic Hall system made with non-relativistic pseudospin-1/2 bosonic particles. By tuning the interaction strength, the system can show how the spin term in the particle current is needed for the local matching of classical drift velocity in the ground state of the non-interacting system, whereas for increasing interparticle interactions the spin-to-orbital current ratio decreases. In either case, since the overall particle velocity changes with the spin term contribution, so does its circulation in a loop, which in turn has physical consequences for its relationship with the Aharonov-Bohm phase shift.
	\end{abstract}
	\maketitle
	
	\section{Introduction}
	Particle current in non-relativistic quantum mechanics presents subtleties when spin is considered. For spin-1/2 particles one has to use the Pauli equation, but this leads to the current density not being well defined. It can only be derived without ambiguity from the Dirac equation \cite{Ohanian1986,Nowakowski1999}, but, done this way, an additional spin term arises in the non-relativistic limit.  An alternative derivation, based on consistency considerations within the non-relativistic system, requires the presence of such spin term in the momentum density with the result that the stress tensor (as part of the energy-momentum tensor) can be symmetrized in order to account for the conservation of total angular momentum (see e.g. Ref. \cite{Takahashi1986}). Although this term has null divergence, hence no contribution to the continuity equation (which might be the reason why it is often neglected) it could still be physically relevant, as is expected to be the case in the quantum Hall effect \cite{Nowakowski1999}.
	
	Another quantity with specific features of the SU(2) internal symmetry is spin current density. Such features have recently been observed in atomic superfluids with internal degrees of freedom (see e.g. Refs. \cite{Farolfi2021,Cominotti2023}). Here, as was predicted for non-relativistic Bose-Einstein condensates (BECs) of pseudospin-1/2 particles \cite{Nikuni2003}, the time evolution of the spin density ${\bf s}=({\hbar}/{2})\chi^\dagger\boldsymbol{\sigma}\chi$ (where $\boldsymbol{\sigma}=\{\sigma_x,\sigma_y,\sigma_z\}$ is the vector of Pauli matrices and $\chi=[\chi_\uparrow\,\chi_\downarrow
	]^T$ is the two-component wave function) involves so-called quantum spin torque \cite{Manchon2019} and allows for the simulation of ferromagnetic materials with bosonic gases \cite{Recati2022}. Previously, similar equations of motion for the spin density had likewise been consistently derived when the non-relativistic limit of the Dirac equation is taken, so demonstrating that the additional spin terms compare well with observed spin currents in condensed-matter spin-Hall systems \cite{An2012}. 
	
	The time evolution of both the particle density and the spin vector follows from the existence of conserved currents, which, according to Noether`s theorem, result from underlying continuous symmetries. While the particle number is always a conserved quantity in non-relativistic systems, and then $\partial_t (\chi^\dagger\chi)+\nabla\cdot {\bf J}=0$, where $ {\bf J}$ is the particle current density, 
	the spin current density is only conserved for constant orbital angular momentum and in the absence of external torques, and then $\partial_t {\bf s}+\partial_j\, {\bf J}_{s,j}=0$, with implicit sum over $j=x,y,z$, where $ {\bf J}_{s,j}$ is the spin current tensor. 
	Although specific spin terms have been derived in the spin current tensor of spinor BECs \cite{Nikuni2003} (see also  Ref. \cite{An2012} for a relativistic approach), and a connection between spin density gradients and the curl of the velocity field has been established \cite{Yukawa2018}, notwithstanding analogous spin terms have not been shown to contribute to the particle current density.

	The main goal of the present work is to show that a spin contribution proportional to the curl of the spin density could be observed in the particle current density of spinor Bose-Einstein condensates. This is illustrated in the ground state of both a non-interacting and an interacting pseudospin-1/2 BEC of electrically neutral particles that, subjected to synthetic gauge fields, simulate the integer quantum Hall effect.  By keeping leading order terms in a derivation that starts from the relativistic equation for a spin-1/2 field, the mentioned contribution is obtained and the classical drift velocity in the ground state of the non-interacting system is matched. Hence, by turning off the electric field the local velocity vanishes. However, when the interparticle interaction increases the spin terms decrease their contribution to the particle current. The modified velocity produces a modified circulation around closed loops, which leads also to a different relationship with the magnetic flux quantization derived from the Aharonov-Bohm phase. 
	 
	These effects could be tested in current experiments with ultracold-gas particles where the build-out of pseudospin-1/2 systems involves two hyperfine states of the same atomic species confined in an optical trap \cite{lamporesi2025two}. This can be the case of condensates of $^{23}$Na or $^{87}$Rb atoms in their corresponding ground state manifolds with total angular momentum (and total spin) eigenstate $|F\rangle=|1\rangle$, in which two spatially miscible hyperfine states whose energy degeneracy is broken by an underlying magnetic field, for instance $|m_F\rangle=|\pm1\rangle$ in sodium as in Ref. \cite{Farolfi2021}, or $|m_F\rangle=|1\rangle, |0\rangle$ in rubidium as in Ref. \cite{Lin2011}, are selected to be macroscopically populated while the rotational properties of the atoms are preserved (see Appendix \ref{sec:App_spin} for an extended discussion). Nevertheless, similar spin contributions to particle currents are expected to be found in generic non-relativistic spinor systems that preserve their internal, spatial rotational properties.

	\section{Theoretical Framework}
	\label{sec:Dirac}
	Within a mean-field framework, the Pauli equation for interacting spin-$1/2$ bosonic particles reads,
	\begin{equation}
		i\hbar\partial_t \,\chi = \frac{(\boldsymbol{\sigma}\cdot{\bf \hat p})^2}{2m}\,\chi+g|\chi|^2\chi,
		\label{eq:Pauli}
	\end{equation}
	where the interparticle interaction strength $g=4\pi\hbar^2 a/m$ is assumed to be proportional to the same scattering length $a$, both for intra- and inter-spin components. This assumption, and in general the modulation of the interaction strength that we consider later, would involve the use of Feshbach resonances in the aforementioned ultracold-gas experiments \cite{Chin2010}.

	Equation \eqref{eq:Pauli} can be obtained in the linear regime, $g= 0$, from the non-relativistic limit of the Dirac equation (see Appendix \ref{sec:app}). There, the Pauli spinor $\chi$ is the \textit{large} component of the Dirac field $[\chi\,\Phi]^T$, whose \textit{small} component $\Phi\approx{\boldsymbol{\sigma}\cdot{\bf \hat p}}\,\chi\,/{2mc}$, where $c$ is the speed of light, provides the relativistic corrections.
    Analogously, starting from the Dirac current density 	\({\bf J}_{\rm Dirac} = c\,\bigl(\chi^\dagger\,\boldsymbol{\sigma}\,\Phi + \Phi^\dagger\,\boldsymbol{\sigma}\,\chi\bigr)\), one obtains the corresponding Pauli current density
	\begin{align}
		{\bf J}
		&\approx \chi^\dagger\,\boldsymbol{\sigma}\,\frac{\boldsymbol{\sigma}\cdot{\bf \hat p}}{2m}\chi
		+ \chi^\dagger\frac{(\boldsymbol{\sigma}\cdot{\bf \hat p})^\dagger}{2m}\boldsymbol{\sigma}\,\chi \nonumber\\
		&= \frac{1}{m}\Re\{\chi^\dagger\,{\bf \hat p}\,\chi\}
		+ \frac{\hbar}{2m}\,\nabla\times(\chi^\dagger\boldsymbol{\sigma}\chi).
		\label{eq:current}
	\end{align}
	This equation resembles closely the so-called Gordon decomposition of the relativistic Dirac current into orbital and spin currents \cite{Ohanian1986}.
	As anticipated, the additional spin term ${\bf J}_\sigma=\nabla\times{\bf s}\,/{m}$ is irrelevant for the continuity equation since $\nabla \cdot (\nabla \times \mathbf{s}) = 0$; while it does not contribute to the particle flux across closed surfaces, it does across open surfaces by adding the circulation of the spin density vector around the surface boundary.  In general, the spin current term does not vanish (see Appendix \ref{app:NoB}), although relevant exceptions include systems that present homogeneous spin density and those where the spin density varies only along the direction of spin polarization.
	Since the derivation of spin terms in the current density has been carried out in the absence of interaction, and depends only on the rotational spin properties of the fields considered, it is not constrained by the spin-statistics features of multiparticle systems. 
	One could alternatively have started from a relativistic equation for a bosonic spinor field to obtain equivalent spin terms, apart from the gyromagnetic factor, in the nonrelativistic limit of the integer spin system, while the interaction between particles (as field excitations) does not remove these effects from the many-particle system \cite{Struyve2004}.
		
	It is useful to rewrite Eq. \eqref{eq:current} in order to show explicitly the role of the phase. To do so, as it is customary (see e.g. \cite{Nikuni2003}), we write a generic state 
	\begin{align}
	\chi(x,y,t)= \sqrt{\rho}\,e^{i\alpha}\binom{e^{-i\phi/2}\cos\frac{\theta}{2}}{e^{i\phi/2}\sin\frac{\theta}{2}},
		\label{eq:state}
	\end{align}
	in terms of four real fields, namely,  the particle density $\rho(\mathbf{r},t)$, the total phase $\alpha(\mathbf{r},t)$, the relative phase $\phi(\mathbf{r},t)=\arg\chi_\downarrow-\arg\chi_\uparrow$, and the spin imbalance angle $\theta(\mathbf{r},t)=\cos^{-1}(|\chi_\uparrow|^2-|\chi_\downarrow|^2)$. In this way, the spin vector and the current density become ${\bf s}=\hbar \rho\,(\sin\theta\cos\phi,\, \sin\theta\sin\phi,\,\cos\theta)/2$ and ${\bf J}=\rho\,{\bf v}$, respectively, where the velocity (that is, in general, the superfluid velocity for the interacting BEC) reads 
	\begin{align}
		{\bf v}= \frac{\hbar}{m}\left(\nabla \alpha-\frac{\cos\theta}{2}\nabla\phi+\frac{\nabla\times{\bf s}}{\hbar\, \rho}\right).
		\label{eq:velocity}
	\end{align}
	As can be seen, due to the spin term, the superfluid velocity depends not only on the phase (as usual) but also on the particle density. 
	Equation \eqref{eq:velocity} defines a velocity field in 3D space, and by means of the Helmholtz decomposition can be written as the sum of a potential term plus a solenoidal field. As can be seen, the spin contributes to both parts of the decomposition. 
	
	{\it Gauge interactions.$-$} For a system of electrically charged particles in the presence of an electromagnetic field, assuming a minimal coupling, Eqs. \eqref{eq:Pauli}, \eqref{eq:current} and \eqref{eq:velocity} remain valid by making the substitutions ${\bf \hat p}\rightarrow \boldsymbol{\hat \pi}={\bf \hat p}-q{\bf A}$, ${i\hbar\partial_t}\rightarrow i\hbar\partial_t-q{V}$, and ${\bf  v}\rightarrow {\bf v}+q{\bf A}/m$, where $\{V,{\bf A}\}$ are the electromagnetic potentials. In this way, the Hamiltonian operator in the nonlinear Pauli equation \eqref{eq:Pauli} generalizes to
	\begin{equation}
		\hat{H}=\frac{({\bf \hat p}-q{\bf A})^2}{2m}-\frac{q\hbar}{2m}\boldsymbol{\sigma}\cdot(\nabla\times\mathbf{A})+qV+{g}|\chi|^2,
		\label{eq:PauliH}
	\end{equation}
	where the second term captures the magnetic interaction due to the spin magnetic dipole. The latter quantity contributes to the system energy density (hence, with opposite sign, to the Lagrangian density) with the term $-\boldsymbol{\mu}_\sigma\cdot\mathbf{B}$, proportional to the magnetic dipole $\boldsymbol{\mu}_\sigma=g_s(q/2m)\,\mathbf{s}$ that includes the famous spin gyromagnetic factor $g_s=2$. 
	
	Alternatively, by means of the the product rule $\nabla\cdot(\mathbf{s}\times\mathbf{A})=\mathbf{A}\cdot(\nabla\times\mathbf{s})-\mathbf{s}\cdot(\nabla\times\mathbf{A})$, after dropping total derivatives, the magnetic energy density due to spin can be written in terms of the spin current density $\mathbf{J}_\sigma$, so that the Lagrangian density that gives rise to the Hamiltonian operator Eq. \eqref{eq:PauliH} reads
	\begin{align}
		\mathcal{L}(\chi)&=i\hbar\chi^\dagger \,\partial_t \,\chi \label{eq:PauliL} \\ &
		  -\left[\frac{|({\bf \hat p}-q{\bf A})\chi|^2}{2m}+q\left(V|\chi|^2-\mathbf{A}\cdot\mathbf{J}_\sigma\right)+\frac{g}{2}|\chi|^4\right], \nonumber
	\end{align}
	which shows explicitly the presence of the spin current term in the coupling with the magnetic field. 
	The total current density can be calculated as the system's response to a gauge field perturbation $\mathbf{J}=(1/q)\delta \mathcal{L}/\delta \mathbf{A}$ (see Ref. \cite{Landau2013} \S 114, and also the related discussion in Ref. \cite{Nowakowski1999}). From the Lagrangian density Eq. \eqref{eq:PauliL} it leads to
		\begin{align}
			\mathbf{J}=\frac{1}{2m}\left[\chi^\dagger(\mathbf{\hat p}-q\mathbf{A})\chi+h.c.\right]+\mathbf{J_\sigma},
		\end{align}
	in agreement with the general form of Eq. \eqref{eq:current}. Nevertheless, notice that the product rule that led to Eq. \eqref{eq:PauliL} can be applied only when the spin density is a spatial 3D vector, while this cannot be done when it proceeds from an internal, non-space-related SU(2) symmetry, and then there is no spin term in the current (see Appendix \ref{sec:App_spin}).  
	
	\subsection{Fluxoid quantization and Aharonov-Bohm phase }
	
	In the presence of magnetic fields, the velocity circulation around closed loops in an electrically charged  superfluid  determines the conditions for quantization of the magnetic flux (see e.g. \cite{Barone1982} Ch. 12). These conditions are more involved in spin-1/2 systems, where, according to Eq. \eqref{eq:velocity}, such circulation $\Gamma=\oint{\bf v}\cdot {\bf d}\boldsymbol\ell=\oint{\bf J}\cdot {\bf d}\boldsymbol\ell/\rho$ results in 
	\begin{align}
		\Phi+\frac{m}{q}\Gamma= \frac{\Phi_0}{2\pi}\oint\left(\nabla \alpha-\frac{\cos\theta}{2}\nabla\phi+\frac{\nabla\times{\bf s}}{\hbar\, \rho}\right)\cdot{\bf d}\boldsymbol\ell,
		\label{eq:circulation}
	\end{align}
	which provides a relationship between magnetic flux $\Phi=\oint \mathbf{A\cdot}{\bf d}\boldsymbol\ell$ and spinor phases. The left hand side of Eq. \eqref{eq:circulation} is the so-called fluxoid $\Phi'$ that in scalar superfluids becomes quantized, in units of the flux quantum $\Phi_0=h/q$, due to the uniqueness condition of the wave function and its dependence on the only phase $\alpha$.  On the contrary, the fluxoid is in general not quantized in spin-1/2 systems because of its dependence on both the population imbalance ($\propto \cos\theta$) and the spin density. This dependence leads also to a distinct relationship (with respect to scalar condensates) for the  Aharanov-Bohm phase $2\pi\Phi/\Phi_0=2\pi(\Phi'-m\Gamma/q)/\Phi_0$ accumulated in a closed loop.

	For later convenience, we define the orbital fluxoid $\Phi_{orb}'=\Phi+m\Gamma_{orb}/q$, where the circulation $\Gamma_{orb}=\oint (\mathbf{J}-\mathbf{J_\sigma})\cdot{\bf d}\boldsymbol\ell/\rho$ is only contributed by the orbital part of the particle current, so that 
		${\Phi_{orb}'}=  (\Phi_0/{2\pi} )\oint(\nabla\alpha-\cos\theta\nabla\phi/2)\cdot{\bf d}\boldsymbol\ell$. In this way
	a fluxoid quantization condition can be restored if the system becomes frozen in a state with fixed imbalance, where
	\begin{align}
		\frac{\Phi_{orb}'}{\Phi_0}= j_\uparrow\,\cos^2\frac{\theta}{2} +j_\downarrow\,\sin^2\frac{\theta}{2}.
		\label{eq:circulation1}
	\end{align}
	and $j_{\uparrow\downarrow}=0,\pm 1,\pm 2, \dots$ are the winding numbers of the spin components.
	Systems with equal winding numbers $j_\uparrow=j_\downarrow=j$ or full imbalance, $\theta=$0 or $\pi$, recover the condition for scalar superfluids ${\Phi_{orb}'}=j\,{\Phi_0}$. 
	On the other hand, systems with no imbalance, $\theta=\pi/2$, can produce either integer or half integer values of the fluxoid  quantization ${\Phi_{orb}'}=(j_\uparrow+j_\downarrow)\,{\Phi_0}/2$.
	For closed loops along which the circulation of the orbital velocity vanishes, $\Gamma_{orb}=0$ and $\Phi=\Phi_{orb}'$, the fluxoid quantization transforms into a flux quantization condition with Aharanov-Bohm phase factor $\exp[i\pi(j_\uparrow+j_\downarrow)]$, which can change the sign of the wave function. The half integer multiples of the flux quantum are made possible by the periodic variation of the relative phase $\phi\rightarrow\phi+2\pi$, as happens in loops encircling Josephson vortices \cite{Barone1982}.

	\section{The synthetic Hall system}
	\label{sec:Hall}
	
	In order to study a Hall system, where the motion takes place within the $x-y$ plane, a constant, perpendicular magnetic field  $B_z$, with $z$-direction, is considered along with an in-plane, constant electric field $-E_x$, with negative $x$-direction (see Appendix \ref{app:coherentcoupling}).
	We write the electromagnetic potentials in the so-called Landau gauge,
	$	V = E_x\,x,\quad {\bf A} = (0,  B_z\,x,0),$ and, as a result, the Pauli Hamiltonian in Eq. \eqref{eq:PauliH} becomes
	\begin{equation}
		\hat H
		= \frac{\hat p_x^2 + (\hat p_y - qB_z x)^2}{2m}
		- \frac{q\hbar}{2m}\,\sigma_z\,B_z+ q\,E_x\,x+g|\chi|^2,
		\label{eq:Hall}
	\end{equation}
	where we have not included the dependence on the $z-$coordinate, assuming that the system is frozen in its ground state along this direction and consequently the interaction strength is renormalized as an equivalent 2D strength (that we will also denote by $g$ from now on). 
	
	Since $[\hat H,\sigma_z]=0$, and $[\hat H,\hat p_y]=0$ if $|\chi|^2$ does not depend on the $y-$coordinate, one can search for solutions to Eq. \eqref{eq:Pauli}  with the functional form
	\begin{equation}
		\chi(x,y,t) = \binom{c_\uparrow}{c_\downarrow}\psi(x)\,\frac{1}{\sqrt{L_y}}e^{i (k y-\mu t/\hbar)},
		\label{eq:sol}
	\end{equation}
	where $L_y$ is the system length along $y$, and $[{c_\uparrow}\,{c_\downarrow}]^T$ is one of the eigenvectors of the $\sigma_z$ operator, either $[1\,0]^T$ or $[0\,1]^T$,  with eigenvalues $\sigma=\pm 1$, respectively, which characterize fully spin-polarized states.

	\begin{figure}
		\centering
		\includegraphics[width=0.95\columnwidth]{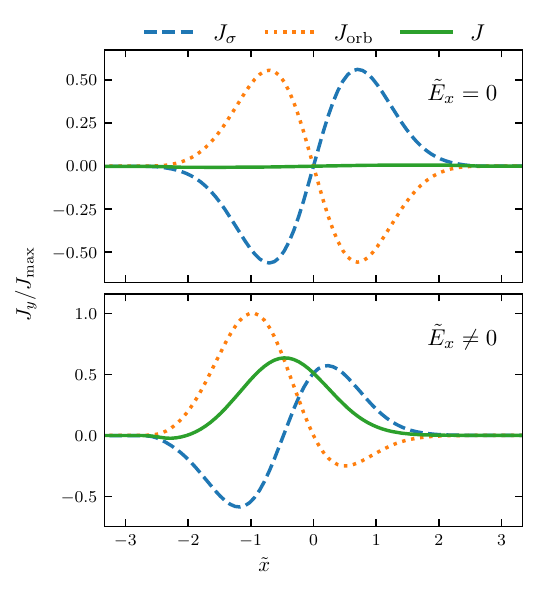}
		\caption{
			Current density in the lowest Landau level ($n=0$) of a system with $E_x=0$ (top) and $\tilde E_x=E_x/B_z= 0.5\,\omega_B\ell_B$ (bottom) as a function of the non-dimensional coordinate $\tilde x=(x-x_k)/\ell_B$.
			As predicted by Eq. \eqref{eq:Jy}, the orbital and spin contributions that are odd in $\tilde x$ cancel locally ($\mathbf{J}=0$) in the absence of electric field, and produce the classical drift velocity $\mathbf{J}/|\chi|^2=0.5\,\omega_B\ell_B$ in the presence of electric field. Units of $J_{max}=4.36\times 10^{-2}\,\omega_B/\ell_B^2$ are used.
		}
		\label{fig:Hallcurrentsspindow}
	\end{figure}
	\subsection{The non-interacting Hall system}
	
	 It is insightful to start from the non-interacting limit, $g=0$, where the analytical solutions to the eigenvalue equation for $\psi(x)$ are known. From the direct substitution of \eqref{eq:sol} in Eq.  \eqref{eq:PauliH}, one gets
	\begin{equation}
		\left[\frac{\hat p_x^2}{2m} + \frac{m\omega_B^2(x - x_k)^2}{2} \right]\psi_k=\left(\mu-\mathcal{E}_k+\sigma\frac{\hbar\omega_B}{2} \right) \psi_k,
		\label{eq:Landau}
	\end{equation}
	where we have defined
	\begin{equation}
		\label{x0:Hall}
		\omega_B = \frac{qB_z}{m},
		\quad
		x_k = \frac{\hbar k}{qB_z} - \frac{mE_x}{qB_z^2}.	\quad \mathcal{E}_k=\frac{\hbar k E_x}{B_z}- \frac{mE_x^2}{2B_z^2},
	\end{equation}
	and assume from now on, without loss of generality, that $q>0$ and $B_z>0$.
	The solutions to Eq. \eqref{eq:Landau} are (real) Hermite functions \cite{Abramowitz1948} $\psi_{k,n}=\mathcal{H}_n[(x-x_k)/\ell_B]\exp[-(x-x_k)^2/2\ell_B^2]/\sqrt{\ell_B}$, product of a normalized Hermite polynomial $\mathcal{H}_n$ of order $n$ times a gaussian of width $\ell_B=\sqrt{\hbar/m\omega_B}$, and give rise to the appearance of the celebrated Landau levels \cite{Landau2013} in the energy spectrum of the Hamiltonian Eq. \eqref{eq:Hall}, 
	\begin{align}	
		\mu_{k,n,\sigma} = \mathcal{E}_k+\hbar\omega_B\Bigl(n+\frac{1-\sigma}{2}\Bigr).
	\end{align}	
	labeled by the excitation number $n=0,1,2,\dots$ of a 1D harmonic oscillator characterized by the cyclotron frequency $\omega_B$. The total eigenenergy depends also on spin, labeled by $\sigma$, and, due to the presence of the electric field, on the momentum $\hbar k$ along the $y-$direction.
	For generic stationary eigenstates $\chi_{k,n,\sigma}(x,y)$ the Pauli probability current Eq. \eqref{eq:current} gives a Hall current
	\begin{equation}
		\label{eq:J}
		{\bf J }= \frac{|\psi_{k,n}|^2}{L_y}\, \left(0,\frac{\hbar k}{m}-\omega_B\,x\,-\sigma\,\frac{\hbar}{2m}\,\partial_x\ln|\psi_{k,n}|^2,0\right),
	\end{equation}
	along the $y-$direction, perpendicular to the applied electric field, with a spin current contribution ${J_y}_\sigma=-\sigma\,{\hbar}\,\partial_x|\psi_{k,n}|^2/{2m}$ of opposite sign for the two spin components. 

	In the lowest Landau level, $n=0$, by using the explicit form of the eigenfunctions, the spin current density becomes
	\begin{equation}
			{J_y}_\sigma =\frac{\sigma\, \psi_{k,0}^2}{L_y}\, (x-x_k)\,{{\omega_B}},
	\end{equation}
	while the total current density is
	\begin{equation}
		{J_y}	=\frac{\psi_{k,0}^2}{L_y}\,\left[(\sigma-1)(x-x_k)\,{{\omega_B}}+\frac{E_x}{B_z}\right],
		\label{eq:Jy}
	\end{equation}
	and the orbital current density is the difference ${J_y}-{J_y}_\sigma$.
	In particular, the spin polarized ($\sigma=1$) eigenstates
	have energy eigenvalues $\mathcal{E}_k$, and current density ${J_y}	=|\chi_{k,0,1}|^2 \,{E_x}/{B_z}$,
	whose characteristic velocity ${J_y}/|\chi|^2$ matches the classical \textit{drift} velocity $v_d={E_x}/{B_z}$. In the absence of electric field, the degenerate ground state of the non-interacting system has no currents (see also Appendix \ref{app:GeneralCancellation}). It is worth noticing that these results are local properties of the spinor system, in contrast with the expectation values of the corresponding currents obtained in a scalar system (see, e.g. Ref. \cite{Tong2016}).
	Figure \ref{fig:Hallcurrentsspindow} represents these results for a system confined in a 2D box, with periodic boundary conditions along the $y-$direction, in two scenarios: with $E_x=0$ (top panel) and ${E_x}/{B_z}= 0.5\,\omega_B\ell_B$ (bottom panel). As can be seen, in the latter case the spin-current term does not vanish, and is odd in the shifted $\tilde x-$coordinate; after adding the orbital term, the total current density shows an even Gaussian profile whose peak is proportional to the applied electric field.

\begin{figure}[t]
	\centering
	\includegraphics[width=\columnwidth]{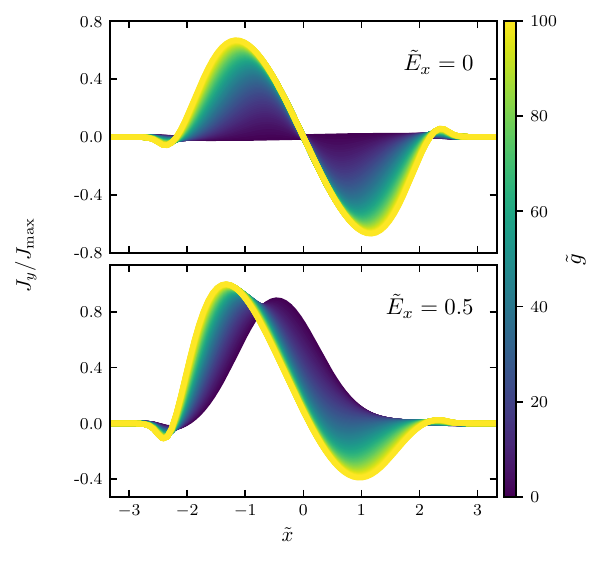}
	\caption{ Transverse profiles of the total current density $J_y(x)$ in repulsively interacting BECs, confined within a 2D box of size $L_x\times L_y=5\ell_B\times10 \ell_B$ (with periodic boundary conditions along the $y-$direction) and subjected to a perpendicular magnetic field $B_z$, for a continuous sweep of the interaction strength $\tilde g=mgN/\hbar^2 \in [0, 100]$. Top: In the absence of electric field, $E_x=0$. Bottom:  For electric field $\tilde E_x=E_x/B_z=0.5\,\omega_B\ell_B$. Units of $J_{max}=3.13\times 10^{-2}\,\omega_B/\ell_B^2$.
	}
	\label{fig:g_sweep}
\end{figure}	
\begin{figure}[t]
	\centering
	\includegraphics[width=0.85\columnwidth]{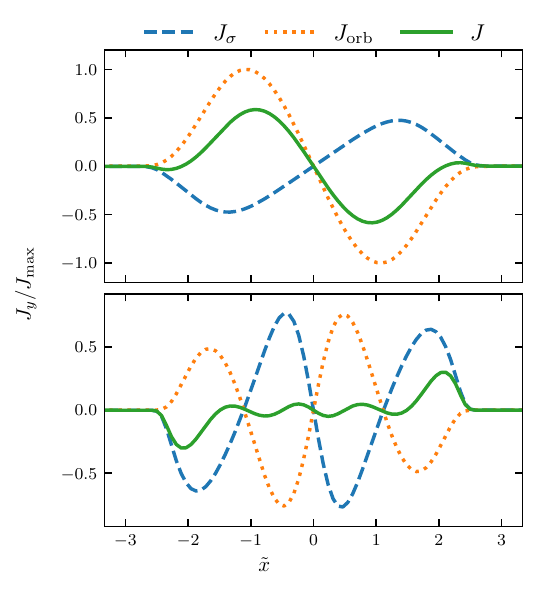}
	\caption{
		Transverse profiles of spin $\mathbf{J}_\sigma$, orbital $\mathbf{J}_{orb}$, and total $\mathbf{J}$ current densities in systems with interaction strength $gN=50\,\hbar^2/m$, and vanishing electric field $E_x=0$. Top: $y-$independent currents in a smooth, excited state without vortices. Bottom: Currents on the line across the core of one of the two vortices that emerge in the ground state of the system.	 Current densities are obtained by means of Eq. \eqref{eq:current} from the numerically computed stationary state evolved with Hamiltonian Eq. \eqref{eq:Hall}. Units of $J_{max}=2.69\times 10^{-2}\,\omega_B/\ell_B^2$ are used.	
	}
	\label{fig:with_out}
\end{figure}
\subsection{The interacting Hall system}

	Further analytical insight can be obtained by considering the regime dominated by interactions, where $\mu\gg \hbar\omega_B$. In this case, the  Thomas-Fermi ansatz 
	\begin{equation}
		\psi_{TF}(x) = \sqrt{\frac{\mu_{eff}-\frac{1}{2}m\omega_B^2(x-x_k)^2}{g/L_y}},
		\label{eq:TF}
	\end{equation}
	for $\psi_{TF} \ge 0$ and $\psi_{TF}=0$ otherwise, where $\mu_{eff}=\mu-\mathcal{E}_k+{\hbar\omega_B}/{2}$, approximates the $x-$dependence of the ground state (with $\sigma=1$) by neglecting the kinetic energy in this direction, along which the condensate extension, $x\in[x_k-R_{TF},\,x_k+R_{TF}]$, is determined by the Thomas-Fermi radius $R_{TF}=\sqrt{2\mu_{eff}/m}\,/\omega_B$. By enforcing normalization to the number of particles $N$ in the condensate, $\int |\chi_{TF}|^2 dxdy=N$, one obtains the chemical potential as a function of the electromagnetic fields,
	\begin{align}
		\mu=\mathcal{E}_k-\frac{\hbar\omega_B}{2}+\left(\frac{3\sqrt{m}\,\omega_B g N}{4\sqrt{2}L_y}\right)^{2/3}.
		\label{eq:mu}
	\end{align}
	
	 In contrast with the linear case, the current density 
	\begin{align}
		J_y=\left(\frac{\hbar\omega_B}{2g}-|\chi_{TF}|^2\right)(x-x_k)\omega_B+|\chi_{TF}|^2\frac{E_x}{B_z},
		\label{eq:Jtf}
	\end{align}
	takes nonzero values in the absence of electric field, due to the fact that the spin current density, associated with the term $\hbar\omega_B/2g$ in the leftmost parenthesis of Eq. \eqref{eq:Jtf}, does not cancel the antisymmetric contribution around $x_k$ of the orbital current. This fact could pose a prospective method to infer the interaction strength $g$ through the measurement of currents in experiments, as far as the non-interacting ground state evolves smoothly, for small $g$, into the interacting regime. 
	
    The Thomas-Fermi ansatz, through Eq. \eqref{eq:Jtf}, demonstrates the presence of high, counter-propagating currents around the center lines $x=x_k$ for increasing values of the interaction or magnetic field (see Fig. \ref{fig:g_sweep}), which, as in classical fluids, can lead to instabilities of the Kelvin-Helmholtz type (see, for instance, \cite{Thorne2017} Ch. 14) and then to vorticity (see \cite{Feynman2018}, Ch. 11). As a result, the emergence of quantum vortices in the ground state is expected beyond a particle current threshold, or alternatively beyond a magnetic field threshold. Such a critical value can be found, for instance, by mapping the system threaded by the magnetic field into a system  that rotates with the Larmor frequency $\Omega_L=\omega_B/2$ (see Appendix \ref{app:critical_field} for an extended discussion), and use the expressions for the critical angular frequency that give rise to vortices. 
    
    \begin{figure*}[htbp]
    	\centering	
    	\includegraphics[width=\linewidth]{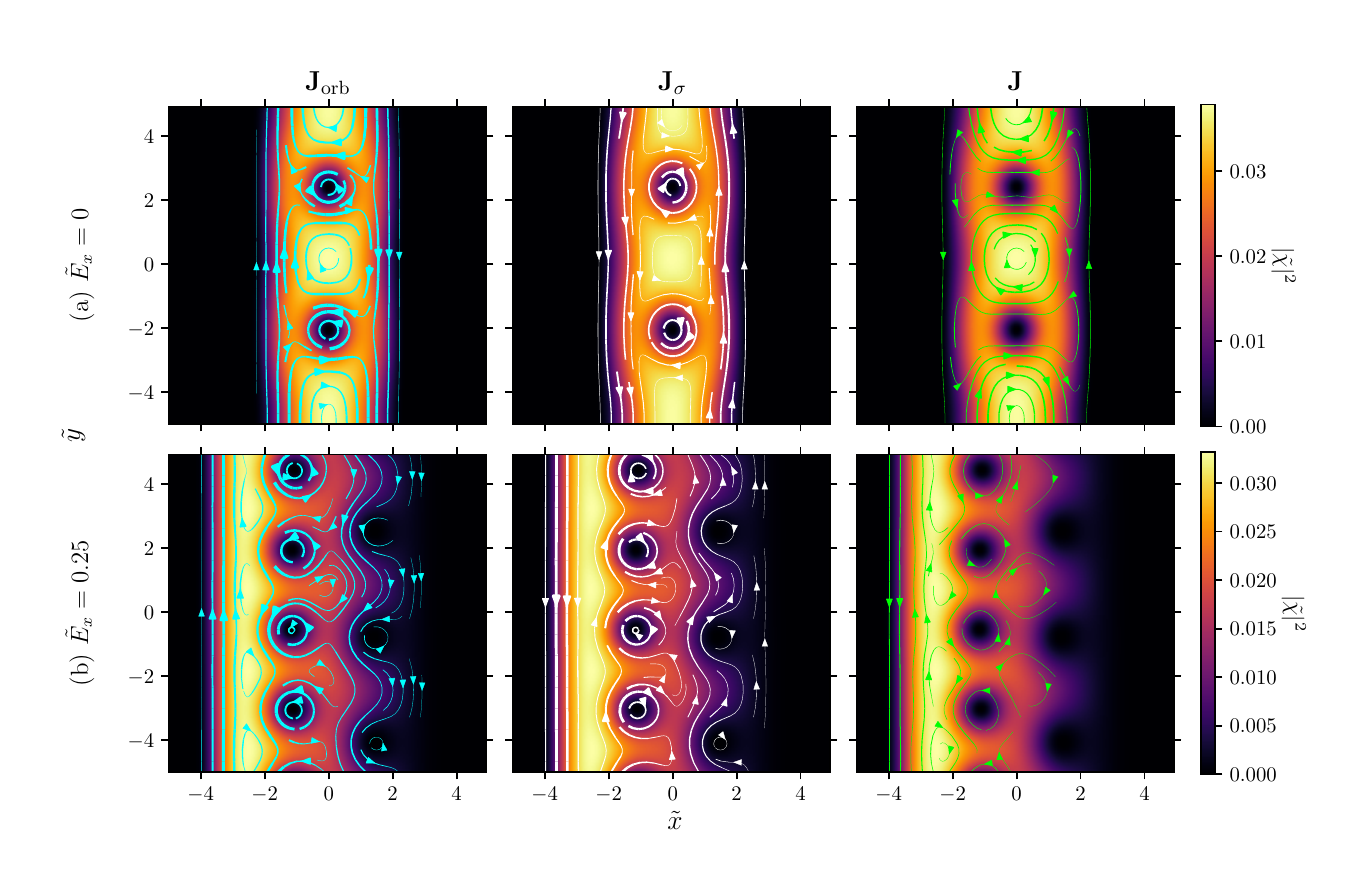}
    	\caption{Spatial distribution of the orbital $\mathbf{J}_{orb}$, spin $\mathbf{J}_\sigma$,  and total $\mathbf{J}$ current densities (represented by streamlines of width proportional to current magnitude) in a BEC with interaction stength $gN=50\,\hbar^2/m$ trapped by box potential and subjected to a perpendicular magnetic field $B_z$. The underlying color maps represent the normalized particle density $|\tilde\chi|^2=(\ell_B|\chi|)^2/N$. (a) In the absence of an external electric field, $E_x=0$ and within a box size $5\ell_B\times10 \ell_B$, the ground state nucleates two vortices. (b) In a wider box $8\ell_B\times10 \ell_B$, and for an electric field $E_x/B_z=0.25\,\omega_B	\ell_B$ a vortex lattice emerges displaced and distorted due to the combined effects of local quantized circulation around vortex cores and overall Hall drift. Current densities were obtained by means of Eq. \eqref{eq:current} from the numerically computed stationary state evolved with Hamiltonian Eq. \eqref{eq:Hall}.}
    	\label{fig:vortices}
    \end{figure*}
	\section{Numerical results}
	\label{sec:numerics} 
	
	We use a 2D numerical domain that can simulate the dynamics of a quasi-two-dimensional BEC loaded in a box potential, as is currently done in ultracold gas experiments (see e.g. \cite{chomaz2015,navon2021}). The box, of area $L_x\times L_y$, is defined in the $x-y$ plane by the combination of the soft-wall potential
		\begin{equation}
		\label{eq:well_potential}
		V(x) = 10\mu \, \left[ 1 +  \tanh\left(\frac{|x| - 0.5\, L_x}{0.2\,\xi}\right)\right],
	\end{equation}
	where $\mu$ and $\xi=\hbar/\sqrt{m\mu}$ are the chemical potential and healing length, respectively, and periodic boundary conditions along the $y$-direction.
	By means of imaginary time evolution, the ground state of the system is found for varying parameters: box size, electric field magnitude, and interaction strength. 
	
	Figure \ref{fig:g_sweep} depicts the transverse profile (along the $x-$direction) of total current densities that are uniform in the $y-$direction for varying interaction strength. They are shown in absence  ($E_x=0$, top panel) and presence ($E_x/B_z=0.5\,\omega_B\ell_B$, bottom panel) of electric field, with fixed magnetic field, and  within a box of size $L_x=5\,\ell_B$ and $L_y=10\,\ell_B$. Both panels gather the numerical results in a range of interaction values $gN \in [0, 100]\,\hbar^2/m$, covering states between the non-interacting and the beginning of the Thomas-Fermi regimes, up to $\mu=2.8\,\hbar\omega_B$. As a general feature, due to the spin current term, small edge currents make their appearance with opposite direction to the bulk currents, and
	both the wave functions and currents are shifted proportionally to the strength of the electric field $\Delta x = -mE_x/qB_z^2$. 

However, as anticipated, the $y-$independent density and current profiles in the ground state change due to the emergence of quantum vortices. This happens, for fixed magnetic field $B_z$, when the value of interaction $gN$ is higher than a critical threshold (see, e.g. \cite{Lin2009, lamporesi2025two}). 
In the systems considered in Fig. \ref{fig:g_sweep}, our results show that the ground states of systems with $gN>13\,\hbar^2/m$ are vortex states, with a minimum of two vortices due to the periodic boundary conditions. 
Therefore, beyond this threshold, the currents represented in Fig \ref{fig:g_sweep} correspond to excited states.
Typical differences in the current-density profiles of vortex and non-vortex states are shown in Fig. \ref{fig:with_out} for a system with $g N=50\,\hbar^2/m$. The top panel shows the $y-$independent spin and orbital currents in the non-vortex phase, which can be approximated  by the analytical expressions \eqref{eq:TF} and \eqref{eq:Jtf}. In contrast, the currents shown in the bottom panel correspond to a vortex state that includes two equidistant vortices along the $x=0$ line [and are also represented with their $x-y$ dependence, on the top of the 2D density map, in Fig. \ref{fig:vortices}(a)] and, prominently, reach much higher values at the edges due to the spin contribution. As before, the spin currents tend to cancel orbital currents, but they are weakened in the bulk for increasing interactions.

The appearance of the vortex phase depends also on the box size, and the critical magnetic field is lower for larger systems (it scales as $\propto\,\ln(R/\xi)/R^2$ in a cylindrical system of radius $R$, see Appendix \ref{app:critical_field}).
For instance, for systems with the same parameters used in Fig. \ref{fig:with_out} but increasing the box width up to $L_x=8\,\ell_B$, the ground state becomes a vortex lattice as soon as $gN>1\,\hbar^2/m$,
	The  presence of electric field has a significant effect in the vortex phase since the vortices have to accomodate in a shifted density profile. As can be seen in the bottom panels of Fig. \ref{fig:vortices}, for a case with $E_x/B_z=0.25\,\omega_B\ell_B$, the varying density profile translates into vortices with different core sizes, due to the non-uniform healing length  $\xi(x) \propto 1/\sqrt{g\rho(x)}$.	The role of the spin term in the current density is still manifest at the edges of the system, due to the gradients of the spin density.

	\section{Conclusions}
	
	 When the local rotational symmetry of the constituent particles is preserved, the current density of spinor BECs includes additional terms proportional to the gradients of the spin density that are often neglected.
	This feature is shared by generic non-relativistic systems whose particles have a spatial spin degree of freedom, as  is the case of the ground state of the hydrogen atom. We have explored the contribution of these current terms in pseudospin-1/2 ultracold atomic gases, where the particle states that are macroscopically populated belong to two of the three energy levels of the same hyperfine manifold. Our analysis relies on the fact that the gyromagnetic $g$-factor is exactly 2, which results from Dirac and Pauli equations (see related discussion in Ref. \cite{Nowakowski1999}).
	
	We have focused on the ground state of the system subjected to synthetic gauge fields in order to simulate the integer quantum Hall effect in ultracold atomic gases;  notwithstanding, the mentioned spin terms do not depend on the presence of gauge fields.
	In non-interacting systems of this type the effect of the spin current is dramatic since it cancels local orbital currents in the absence of electric fields, and results in the classical Hall drift when electric fields are present. 
	In BECs that interact repulsively, spin current opposes, but does not cancel, the orbital current; in this case, the relevant contribution of the spin term manifests as edge currents, even when the system enters the vortex phase.

	Although we have chosen a particular case of confinement and have not distinguished between the interaction strengths of particles belonging to different spin components, which is something that we aim to explore in future work, we expect that these generic spin effects in the particle current of finite systems could be measured in ultracold-gas setups (assuming that Feshbach resonances are available to modulate interactions), like those employed in the observation of similar spin-related properties in atomic superfluids \cite{Farolfi2021,Cominotti2023}.
	Since the additional spin terms in the current contribute to the mechanical momentum of the particles (notice that the momentum density is $m$ times the current density \cite{Takahashi1986}), but do not rely on phase gradients that could be observed with interference measurements, it is expected that usual absorption imaging after ballistic expansion of the spinor condensate can be able to measure the contributions of these terms in the total in-trap momentum of the atomic cloud. For this purpose, the best scenario to verify the theoretical predictions would correspond to spinor states whose current density consists of spin terms only, that is $\mathbf{J}=\mathbf{J_\sigma}$; among these, for instance, a confined, fully spin polarized state in the absence of gauge fields, defined by a constant phase and a non-homogeneous density profile, would show in general a non-vanishing, divergence-free current density which is absent in an equivalent scalar condensate. Additional measurements that use nondestructive imaging of the condensate by means of phase-contrast imaging or polarized light could help determine quantitatively the local values of particle currents and spin density \cite{Higbie2005,Vengalattore2008}.

	\section*{Acknowledgements}
	
	We are grateful to the anonymous referees that encouraged us to discuss the origin, and clarify the experimental conditions for possible observations, of the considered spin effects.
	
	 \bibliographystyle{unsrt} 
	 \bibliography{spin_bib}
	
	\appendix
	\section{Considerations on pseudospin-1/2 condensates}
	\label{sec:App_spin}
	
	The constituent particle states considered in this work are meant to correspond to two hyperfine levels of the same atomic species and isotope. As usual in ultracold-gas experiments with $^{23}$Na or $^{87}$Rb atoms, these states could be part of the ground-state, hyperfine $|F=1\rangle$ manifold that in an optical trap would present energy degeneracy for the three sublevels $m_F=1,0.-1$. 
	
	Under these conditions, without making yet the selection of two particular, out of the three, hyperfine states to generate a fragmented two-component condensate, the rotational spin symmetry of the manifold, furnishing the fundamental three-dimensional (spin-1) vector representation of the SO(3) symmetry group,  is expected to be manifest in a spin term within the current density with the same functional form as written in Eq. \eqref{eq:current}, but substituting the Pauli matrices by corresponding spin-1 rotation generators $S_j$ (or $3\times 3$ Pauli matrices).  This consideration is supported not only by relativistic arguments similar to those presented in Sec. \ref{sec:Dirac} (see Ref. \cite{Struyve2004}), but also from consistency in the definition of the conserved total angular momentum in the non-relativistic system that requires an spin term to symmetrize the energy-momentum tensor (see Ref. \cite{Takahashi1986}). 
	This term endows the velocity field of the spinor with additional rotational features (see Ref. \cite{Yukawa2018} for a generalization of the Mermin-Ho relation in this system) where  the curl of the velocity involves gradients of the spin density, as built in this case  with $3\times3$ spin-rotation Pauli matrices.
	Spin interconversion through two-particle scattering is not necessary for such terms to be part of the current density, for they arise from an intrinsic property of the constituent particles, irrespective of the population of the spin components.

	The discussion is more intricate when the process of making the two-component BEC is considered. To this end, in the presence of a bias magnetic field, by means of an emergent quadratic-Zeeman energy term,  one of the hyperfine states in the triplet can be energetically shifted away from the other two, which facilitates the formation of an effective two-component, pseudospin 1/2 BEC,  as realized, for instance, in Ref. \cite{Lin2011}. The resulting system is described by macroscopically populated single-particle states that belong to a subspace of the larger vector space of spin-1 single-particle states, where the local rotational properties are preserved. 
	
	On the other hand,
	in the resulting two-component BEC one can identify, as usual, an additional SU(2) internal symmetry whose transformations are represented by internal 2$\times$2 Pauli matrices not related to spatial rotations, hence the pseudospin 1/2 character of the BEC. The presence of coherent coupling, with explicit interconversion between the two populated spin components enforced by either radiofrequency or Raman-laser radiation, or the implementation of artificial gauge potentials with laser radiation that can modify the orbital currents, act on the system's state as operators that belong to the group of internal SU(2) transformations, and can be interpreted as the interaction of the system with synthetic electromagnetic fields.
	
	 We assume that the two mentioned, different symmetries, spatial SO(3) and internal SU(2), coexist in the two-component BEC and endow it with different properties, while only the rotational symmetry is responsible for the emergence of spin terms in the particle current density.

	\section{From Dirac to Pauli field equations}
	\label{sec:app}
	
	The Dirac Hamiltonian takes the compact form  \cite{Weinberg1995}
	\begin{equation}
		\hat H_{\rm Dirac} = c\,\boldsymbol{\alpha}\cdot{\bf \hat p} + \beta\,m c^2,
		\label{eq:Dirac}
	\end{equation}
	in the representation where 
	\begin{equation}
		\boldsymbol{\alpha}=\begin{pmatrix}0 & \boldsymbol{\sigma}\\ \boldsymbol{\sigma} & 0\end{pmatrix}.
		\qquad
		\beta=\begin{pmatrix}\hat{\mathds{1}}_{2\times2} & 0\\ 0 & -\hat{\mathds{1}}_{2\times2}\end{pmatrix},
	\end{equation}
	From Eq. \eqref{eq:Dirac}, and by writing the Dirac field in terms of its two-component spinors $[\chi\,\Phi]^T$, one obtains the two coupled equations of motion
	\begin{align}
		i\hbar\,\partial_t\chi &= c\,\boldsymbol{\sigma}\cdot{\bf \hat p}\,\Phi + mc^2\,\chi, \label{D1} \\
		i\hbar\,\partial_t\Phi &= c\,\boldsymbol{\sigma}\cdot{\bf \hat p}\,\chi - mc^2\,\Phi,  \label{D2}
	\end{align}
	and the current density	\(\vec J_{\rm Dirac} = c\,\bigl(\chi^\dagger\,\boldsymbol{\sigma}\,\Phi + \Phi^\dagger\,\boldsymbol{\sigma}\,\chi\bigr)\).
	Assuming energies near the rest mass
	\(\mathcal E_{total}=\mathcal E+mc^2\) with \(\mathcal E\ll mc^2\), and solving in Eq. \eqref{D2} for the \textit{small} component
	\begin{equation}
		\label{spinor2}
		\Phi = \frac{c\,\boldsymbol{\sigma}\cdot{\bf \hat p}}{E+mc^2}\,\chi
		\approx \frac{\boldsymbol{\sigma}\cdot{\bf \hat p}}{2mc}\,\chi.
	\end{equation}
	one arrives, in the non-relativistic limit, at the time-independent Pauli equation for the \textit{large} component
	\begin{equation}
		\mathcal E\,\chi = \frac{(\boldsymbol{\sigma}\cdot{\bf \hat p})^2}{2m}\,\chi,
	\end{equation}
	where $\hat H=(\boldsymbol{\sigma}\cdot{\bf \hat p})^2/{2m}$ is the Pauli Hamiltonian. 
	
	\section{Spin currents in the absence of electromagnetic fields}
	\label{app:NoB}
	
	The spin term,  ${\bf J}_\sigma=\nabla\times{\bf s}\,/{m}$ in the current density, does not depend on couplings to external fields, and manifests as an intrinsic property of the spinor system. 
	 As we anticipated, apart from relativistic arguments, one can derive this spin term considering the transformation properties of the spinor field alone against 3D spatial rotations. Following Ref. \cite{Takahashi1986}, one can define the system total angular momentum density for a field $\chi$ with generic spin as
	\begin{align}
		L_j=\epsilon_{jkl} \left(x_k\,T_{l0}+\frac{\hbar}{2}\chi^\dagger S_{kl}\chi\right),
		\label{eq:L}
	\end{align}
	where $\epsilon_{jkl}$ is the totally antisymmetric tensor of order three, $T_{l0}=m\,({J}_{orb})_{l}$ is the $l-$th component of the momentum density as obtained from the system's Lagrangian, and $S_{kl}$ is the corresponding generator of spin rotations around the spatial $j-$axis, which for spin-1/2 reads $S_{kl}=i\epsilon_{klj}\sigma_j/2$.
	In order for the defined total angular momentum to be a conserved quantity in the absence of external torques, the stress tensor (as part of the energy-momentum tensor) has to be symmetrized \cite{Takahashi1986} to be consistent with a momentum density that includes a spin term capable of accounting for Eq. \eqref{eq:L}, namely,
	\begin{equation}
	T_{l0}^s=T_{l0}+i\frac{g_s\hbar}{2}\partial_k\left( \chi^\dagger S_{kl}\chi\right),
	\end{equation}
	which matches $m$ times the total current density, as defined in Eq. \eqref{eq:current},  $T_{l0}^s=m\,[(J_{orb})_l+(J_\sigma)_l]$, with $g_s$ being the corresponding spin gyromagnetic factor, and so $L_j=m\,\epsilon_{jkl}\,x_j[(J_{orb})_l+(J_\sigma)_l/g_s]$.

	For the case study considered in this work,  in the absence of electromagnetic fields, the polarized states with $\mathbf{s}=(0,0,s_z)$ considered in Section \ref{sec:numerics}, confined in a box by a soft-wall potential, still present a non-vanishing current density due entirely to the spin term $\mathbf{J}=\mathbf{J}_\sigma=(0,\,-\partial_x s_z,\,0)/m$. This current density appears associated with the gradients of the spin density that originate at the edges of the system; hence, it remains localized in a region with the typical size of the healing length $\Delta x\approx \xi$. 
	
	The spin term $\nabla\times{\bf s}\,/{m}$ is expected to be also relevant in the time evolution of other quantities in spinor systems that retain rotational symmetry. The continuity equation for the spin density $\mathbf{s}$ presents a significant example, where one can look for generic expressions for the spin current tensor $\mathbf{J}_{s,j}$ with additional terms, other than the quantum spin-torque, associated with $\mathbf{J}_\sigma$ (see, e.g., Ref. \cite{An2012}, in particular, Eq. 15).
	
	\section{Synthetic gauge interactions}
	\label{app:coherentcoupling}
	
		The gauge field terms considered in the case study are assumed to be of synthetic origin \cite{Dalibard2011}. In this way,
		the interaction with the magnetic field is represented in Sect. \ref{sec:Hall}, Eq. \eqref{eq:Hall}, by a Zeeman term, $\propto\sigma_z$, just for convenience of our mathematical model, in order to produce a ground state that is fully polarized. Because of the homogeneous interparticle interaction strength, the unitary transformation $\psi'=U\psi$, with $U=(\sigma_x+\sigma_z)/\sqrt{2}$, can map this model into an equivalent system whose interaction with the magnetic field, due to the rotation $\sigma_x=U\sigma_zU^\dagger$, turns out to have the functional form of the coherent coupling used in ultracold-gas experiments, $\propto\sigma_x$, which produces the interconversion between spin components. To replicate the features of a real Hall system with synthetic electromagnetic fields \cite{Dalibard2011}, a proper gauge potential $\mathbf{A}$, such that $\nabla \times\mathbf{A}=\mathbf{B}$, would have to be also implemented,  for instance, by means of optical fields capable of producing a Berry connection, along the lines of Ref. \cite{Lin2009}, but with the caveat of inducing equal connections in both spin components.

	\section{Current density in the Lowest Landau Level}
	\label{app:GeneralCancellation}
	
	The exact cancellation of the total bulk current $\mathbf{J} = \mathbf{J}_{orb} + \mathbf{J}_{\sigma} = \mathbf{0}$ for the ground state of a noninteracting Hall system in the absence of electric field was explicitly proven for a particular gauge in the specific geometry  of a linear strip. However, such a cancellation constitutes a generic feature for arbitrary states in the Lowest Landau Level (LLL).
	
	In the presence of a magnetic field, the kinetic momentum operators are defined as $\hat{\boldsymbol{\Pi}} = \hat{\mathbf{p}} - q\mathbf{A}$. Following the standard algebraic treatment of Landau levels \cite{Tong2016}, the LLL is characterized as the kernel of the annihilation operator $\hat{a} \propto \hat{\Pi}_x + i\hat{\Pi}_y$, satisfying $\hat{a}\chi = 0$ for any wave function $\chi$ in the degenerate ground state, that is
	\begin{equation}
		(\partial_x  + i\partial_y) \chi = i \frac{q}{\hbar}(A_x+iA_y)\chi.
		\label{eq:a}
	\end{equation}
	We now evaluate the components of the total current for a fully polarized state ($\sigma = 1$),
	with $x$ and $y$ components of orbital and spin currents given by
	\begin{align}
		J_{orb, x} &=  \frac{\hbar}{m}\text{Im}\{\chi^{\dagger} \partial_x \chi\}-\chi^{\dagger}\frac{qA_x}{m}\chi, \\
		J_{orb, y} &= \frac{\hbar}{m}\text{Im}\{\chi^{\dagger} \partial_y \chi\}-\chi^{\dagger}\frac{qA_y}{m}\chi,  \\
		J_{\sigma, x} &= \frac{\hbar}{m}\text{Re}\{\chi^{\dagger} \partial_y \chi\},\\
		J_{\sigma, y} &= -\frac{\hbar}{m}\text{Re}\{\chi^{\dagger}\partial_x \chi\}.
	\end{align}
	Summing up both contributions per coordinate
	\begin{align}
		J_{x} = \frac{\hbar}{m} \text{Im}\{\chi^{\dagger} (\partial_x + i\partial_y) \chi\}-\chi^{\dagger}\frac{qA_x}{m}\chi,\\
		J_{y} = -\frac{\hbar}{m}\text{Re}\{\chi^{\dagger} (\partial_x + i\partial_y)\chi\} -\chi^{\dagger}\frac{qA_y}{m}\chi,
	\end{align}
	and composing the complex current $\mathbf{J}^C=J_{x}+i J_{y}$, one obtains, by means of Eq. \eqref{eq:a},
	\begin{align}
		 \mathbf{J}^C= \frac{\hbar}{im}\left\lbrace\chi^{\dagger} \left[(\partial_x + i\partial_y)-i\frac{q}{\hbar}(A_x+iA_y)\right] \chi\right\rbrace
		 = 0
	\end{align}	
    and, therefore $\mathbf{J} = 0$, which shows the exact cancellation of orbital and spin currents for arbitrary states $\chi$ in the Lowest Landau Level.

	\vspace{0.2cm}
	\section{Critical Field $B_c$ for Vortex Nucleation}
	\label{app:critical_field}
	
	The energy $E'$ of a BEC rotating with angular frequency $\Omega$, as given in the corotating frame of reference, reads (see, e.g. Ref.~\cite{pethick2008bose}) $E' = E - {\Omega}\, L$,
	where $E$ is the energy in the laboratory frame and ${L}$ is the mean value of the angular momentum around the rotation axis. The critical angular velocity $\Omega_c$ for vortex nucleation, so that for $\Omega > \Omega_c$ a vortex state becomes energetically favorable, is
		$\Omega_c = ({E'_L - E_0})/{L}$,
	where $E_0$ is the energy of the vortex-free state and $E_L$ is the energy of the vortex state carrying angular momentum $L$.  For a system trapped in a cylindrical container of radius $R$ this value can be approximated by~\cite{Pitaevskii}
	\begin{align}
		\Omega_{c}
		= \frac{\hbar}{m R^2}
		\ln\left(1.46 \frac{R}{\xi}\right),
		\label{eq:omegaC}
	\end{align} 
	where $	\xi = {\hbar}/\sqrt{2m\mu}$  is the healing length, 
	written in terms of the chemical potential $\mu$.

	This result for rotating condensates can also be used for a BEC in the presence of a synthetic, constant magnetic field. To this end, assuming that the magnetic field points along the $z$ axis, it is convenient to work with the symmetric gauge, $\mathbf{A}=\mathbf{B}\times\mathbf{r}/2$, to obtain the Hamiltonian 
	\begin{equation}
		\label{eq:Hsymmetricgauge}
		\hat{H} = -\frac{\hbar^2}{2m}\nabla^2  + \frac{m\Omega_L^2r^2}{2} - {\Omega_L}\hat{L}_z
	\end{equation} \\
	where the effective angular velocity $\Omega_L = {\omega_B}/{2}=qB_z/(2m)$ can be identified as the Larmor frequency.
	 Therefore,  according to Eq. \eqref{eq:omegaC}, an estimate for the critical magnetic field is
	 $	B_{c} =  {2\hbar}/({q R^2})\,\ln(1.46 {R}/{\xi})$,
	 or in terms of a critical magnetic flux threading the system
		\begin{equation}
		\label{eq:fluxC}
		{\Phi}_c = \Phi_0\,\ln\left(1.46 \frac{R}{\xi}\right).
	\end{equation}

\end{document}